\documentclass[12pt]{article}
\usepackage{jeosman}
\usepackage{epsfig}
\usepackage{amsmath}
\usepackage{amssymb}
\usepackage{mathrsfs}

\newcommand{\rrr}{\boldsymbol{r}}
\newcommand{\GGG}{\boldsymbol{G}}
\newcommand{\EEE}{\boldsymbol{E}}
\newcommand{\DDD}{\boldsymbol{D}}
\newcommand{\aaa}{\boldsymbol{\alpha}}
\newcommand{\kkk}{\boldsymbol{\kappa}}
\newcommand{\vvv}{\boldsymbol{v}}
\newcommand{\nablabf}{\boldsymbol{\nabla}}

\begin{document}

\title{Liquid-infiltrated photonic crystals:\\ Ohmic dissipation and broadening of modes}

\author{Niels Asger Mortensen}

\address{MIC -- Department of Micro and
Nanotechnology, NanoDTU, Technical University of Denmark, DTU-building 345 east, DK-2800 Kongens Lyngby, Denmark}

\email{nam@mic.dtu.dk}

\author{Simon Ejsing}

\address{MIC -- Department of Micro and
Nanotechnology, NanoDTU, Technical University of Denmark, DTU-building 345 east, DK-2800 Kongens Lyngby, Denmark}

\author{Sanshui Xiao}

\address{MIC -- Department of Micro and
Nanotechnology, NanoDTU, Technical University of Denmark, DTU-building 345 east, DK-2800 Kongens Lyngby, Denmark}

\email{Sanshui.Xiao@mic.dtu.dk}


\begin{abstract}
The pronounced light-matter interactions in photonic crystals make
them interesting as opto-fludic "building blocks" for
lab-on-a-chip applications. We show how conducting electrolytes
cause dissipation and smearing of the density-of-states, thus
altering decay dynamics of excited bio-molecules dissolved in the
electrolyte. Likewise, we find spatial damping of propagating
modes, of the order dB/cm, for naturally occurring electrolytes
such as drinking water or physiological salt water.
\end{abstract}

\maketitle

\section{Introduction}

With the emerging field of opto-fluidics~\cite{Psaltis:2006} there
is an increasing attention to liquid-infiltrated photonic crystals
and the development has to a large extend been powered by their
potential use as bio-chemical
sensors~\cite{Loncar:2003,Chow:2004,Domachuk:2004,Kurt:2005,Adams:2005a,Erickson:2006,Hasek:2006,Xiao:2006b}.
Generally speaking, photonic crystals are strongly dispersive
artificial materials, first suggested in 1987 by
Yablonovitch~\cite{Yab:1987} and John~\cite{John:1987}, where a
periodic modulation of the dielectric function causes strong
light-matter interactions for electromagnetic radiation with a
wavelength comparable to the periodicity of the material. Typically,
photonic crystals are made from a high-index dielectric with a
periodic arrangement of voids. Alternatively, free-standing
high-index dielectric structures may be utilized.

From a sensing point of view, light is an often utilized probe in
analytical chemistry and for miniaturized lab-on-a-chip
implementations, combining
microfluidics~\cite{Squires:05,Whitesides:2006} and optics, there is
a strong call for enhanced light-matter interactions compensating
for the reduced optical path length. Liquid-infiltrated photonic
crystals are obvious candidates for this where e.g.
Beer--Lambert--Bouguer absorbance cells may benefit from slow-light
phenomena~\cite{MortensenBLB}. Alternatively, weak perturbations in
the liquid refractive index may cause considerable shifts in
electromagnetic modes, thus making liquid-infiltrated photonic
crystals promising candidates for refractometry~\cite{Xiao:2006b}.

Most bio-chemistry quite naturally occurs in the liquid phase and
typically the host environment is a conducting electrolyte. In
fact, even highly purified water will be weakly conducting due to
the dissociation of water molecules (H$_2$O) into hydrogen (H$^+$)
and hydroxide (OH$^{-}$) ions. In this paper we consider the
effect of Ohmic dissipation and broadening of levels in liquid
infiltrated photonic crystals.

\section{Theory}

We first consider the general case of a photonic crystal where the
unit-cell of the periodic structure is composed of a solid
high-index dielectric (d) material surrounded or infiltrated by a
liquid (l). The corresponding relative dielectric function is given
by

\begin{equation}\label{eq:epsilon}
\varepsilon(\rrr)=\left\{\begin{matrix}\varepsilon_d &,& \rrr\in
{\mathscr V}_{d},\\\\ \varepsilon_l+ i\frac{\sigma}{\omega}&,&\rrr
\in {\mathscr V}_{l}.
\end{matrix}\right.
\end{equation}
where the complex dielectric function of the liquid is quite
similar to the Drude model often employed for the response of
metals at optical frequencies. For liquids it is an adequate
description of bulk properties, thus neglecting possible surface
chemistry and Debye-layer ion accumulation at the interfaces to
the high-index material. Electro-hydrodynamics where momentum is
transferred from the ionic motion to the fluid (see e.g.
Ref.~\cite{Mortensen:2005} and references therein) is also
strongly suppressed for optical frequencies, since the Debye
screening layer forms too slowly in respond to the rapidly varying
optical field. It is common to introduce the Debye response time
$\tau_D=\omega_D^{-1}=\varepsilon_l/\sigma$ and for typical
electrolytes, $\tau_D$ is less than a micro second corresponding
to a Debye frequency in the megahertz regime. For optical
frequencies, $\omega\gg \omega_D$, it is thus fully adequate to
treat the imaginary part in Eq.~(\ref{eq:epsilon}) perturbatively.
The unperturbed electromagnetic modes are governed by the
following generalized eigenvalue problem for the electrical field
\begin{equation}\label{eq:E}
\nablabf\times\nablabf\times\big|\EEE_m\big>=\epsilon\frac{\omega_m^2}{c^2}\big|\EEE_m\big>
\end{equation}
where $\omega_m$ is the eigenfrequency of the $m$th mode, $c$ is
the speed of light in vacuum, and
$\epsilon(\rrr)=\lim_{\sigma\rightarrow 0}\varepsilon(\rrr)$ is
the unperturbed dielectric function characterizing the
electromagnetic problem in the absence of conduction. In the
following the unperturbed eigenmodes are normalized according to
$\big<\EEE_n\big|\epsilon\big|\EEE_m\big>=\delta_{nm}$ where
$\delta_{nm}$ is the Kroenecker delta. From standard first-order
perturbation theory, the effect of a finite
conductivity leads to an imaginary shift in the frequency, %
\begin{equation}
\Delta\omega_m=-\frac{\omega_m}{2}
\frac{\big<\EEE_m\big|i\frac{\sigma}{\omega_m}\big|\EEE_m
\big>_{{\mathscr
V}_l}}{\big<\EEE_m\big|\epsilon\big|\EEE_m\big>_{{\mathscr
V}_{l+d}}}
\end{equation}
where the integral in denominator is restricted to the liquid
region ${\mathscr V}_{l}$ while the integral in the nominator is
over all space, i.e. ${\mathscr V}_{l+d}={\mathscr
V}_{l}+{\mathscr V}_{d}$. Introducing the displacement field
$\big|\DDD_m\big> = \epsilon\big|\EEE_m\big>$ we may rewrite the
result as
\begin{equation}\label{eq:domega}
\Delta\omega_m=-\frac{i}{2}\: \omega_D\times f_m,\quad f_m\equiv
\frac{\big<\EEE_m\big|\DDD_m\big>_{{\mathscr
V}_l}}{\big<\EEE_m\big|\DDD_m\big>_{{\mathscr V}_{l+d}}}
\end{equation}
where $f_m$ is the fraction of dielectric energy localized in the
liquid. Obviously, $f_m$ is a key parameter for refractometry
applications of liquid-infiltrated photonic crystals. For
void-like structures in the evanescent field sensing limit, $f$
will be of the order of a few percent, while for pillar-like
structures the optical overlap with the liquid can be larger than
$50\%$, thus facilitating a much higher sensitivity to refractive
index changes in the liquid~\cite{Xiao:2006b}. For
Beer--Lambert--Bouguer absorbance cells, the slow-light
enhancement also quite naturally scales with
$f$~\cite{MortensenBLB}.

In this paper we will explore two consequences of the small
imaginary shift in the frequency caused by the small imaginary Ohmic
term in Eq.~(\ref{eq:epsilon}). One first obvious consequence is of
course the Ohmic damping. Mathematically, the imaginary shift in
frequency may via the chain-rule be transformed into a small
imaginary shift $\Delta\kkk$ in the Bloch wave vector $\kkk$. The
corresponding damping parameter $\aaa=2\Delta\kkk$ of the Bloch
modes then becomes
\begin{equation}\label{eq:alpha}
\aaa=\frac{f\omega_D}{\vvv_g}
\end{equation}
where $\vvv_g=\partial\omega/\partial\kkk$ is the unperturbed
group velocity. The attenuation will thus quite intuitively
increase with a slowing down of the electromagnetic mode near
photonic-band edges. Modes with a large optical overlap with the
liquid of course suffer most from Ohmic dissipation as reflected
by the proportionality of $\aaa$ to $f$.

Another consequence of the Ohmic conduction is reflected in the
photonic density of states. Quite intuitively, dissipation is linked
to a life-time broadening of the electromagnetic states and in the
density of states this leads to a smearing of sharp features and
even an induced density of states in the gaps where no states exist
in the absence of conduction. To see this explicitly we start from
the following definition of the density of states
\begin{equation}
\rho(\omega)\propto \int_{{\mathscr V}_{l+d}} dr\, {\rm
Im}\{\omega\:{\rm Tr}\, \GGG(r,r,\omega)\}
\end{equation}
where $\GGG$ is the Green's tensor~\cite{Martin:1995,Martin:1999}
defined in accordance with Eq.~(\ref{eq:E}) and the trace is to
sum over the three directions. The density of states is of
particular interest to dipole radiation from e.g. excited
bio-molecules dissolved in the liquid where the decay rate of the
excited state of the molecule is proportional to the
electromagnetic density of states. From the Green's tensor $\GGG$
we get the standard result
\begin{equation}
\rho(\omega)=\frac{1}{{\mathscr V}_{\rm BZ}}\sum_m \int_{{\rm BZ}}
d\kkk\,\frac{2}{\pi}{\rm
Im}\left\{\frac{\omega}{\omega^2-\omega_m^2(\kkk)+i\eta^+
}\right\}
\end{equation}
with an infinitesimal broadening by $\eta^+$ of the
electromagnetic levels. The normalization is given by the volume
${\mathscr V}_{\rm  BZ}= \int_{{\rm BZ}} d\kkk$ of the first
Brillouin zone. For a vanishing $\sigma$ the eigenfrequencies
$\omega_m$ are real and we get the well-known expression
\begin{equation}
\rho_0(\omega)=\frac{1}{{\mathscr V}_{\rm BZ}}\sum_m \int_{{\rm
BZ}} d\kkk\,\delta\left[\omega-\omega_m(\kkk)\right]
\end{equation}
where $\delta(x)$ is the Dirac delta function. For a small but
finite $\sigma$, first-order perturbation theory in the frequency
squared (first order in $\omega_D$) gives
$\Delta(\omega_m^2)=-i\omega_m\omega_Df$ so that
\begin{equation}
\rho(\omega)=\frac{1}{{\mathscr V}_{\rm BZ}}\sum_m \int_{{\rm BZ}}
d\kkk\,\frac{2}{\pi}{\rm
Im}\left\{\frac{\omega}{\omega^2-\omega_m^2(\kkk)+i
\omega_D\omega_m(\kkk)f_m(\kkk) }\right\}
\end{equation}
corresponding to a finite broadening of the order $\omega_D$. Note
how some states are broadened more than others depending on the
filling fraction and eigenfrequency.

Along with the broadening there will be an induced density of
states $\rho_{\rm \scriptscriptstyle PBG}$ in the band-gaps where
no states were available in the absence of conduction. To see this
we consider the center of a band-gap of width $\Delta\Omega$. The
induced density-of-states originates mainly from the upper and
lower band edges with the frequency and the filling factor
evaluated at the edge of the Brillouin zone. From the
corresponding tails we get the following for the induced density
of states
\begin{align}
\rho_{\rm \scriptscriptstyle PBG}\propto
\frac{\omega_D}{(\Delta\Omega)^2 }
\end{align}
demonstrating the competition between the two frequency scales
given by the Debye frequency and the with of the band gap.

\section{Numerical results}

\begin{figure}[t!]
\begin{center}
\epsfig{file=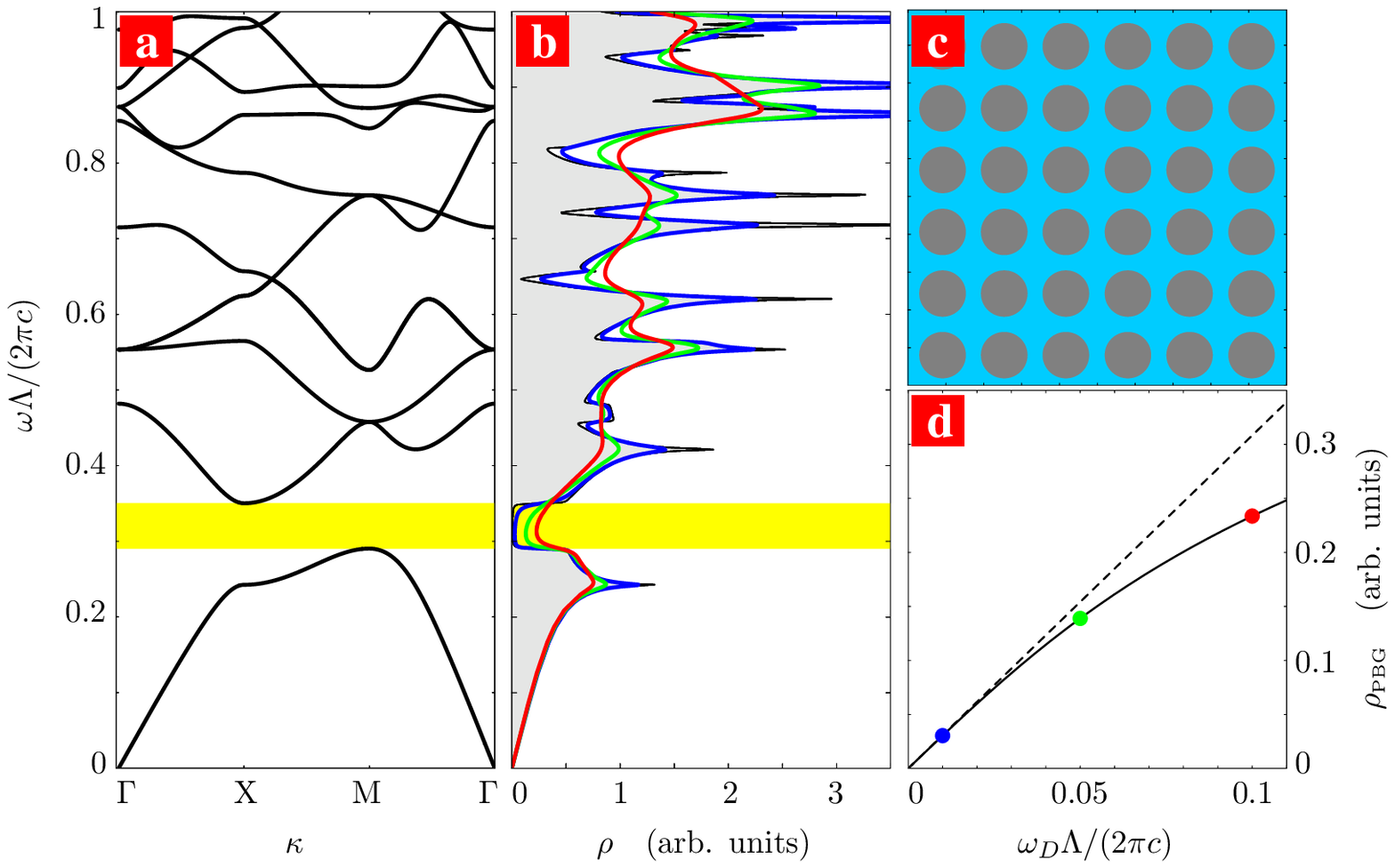, width=1\columnwidth,clip}
\end{center}
\caption{Liquid infiltrated photonic crystal, see Panel (c), with
a square lattice of high-index rods with diameter $d/\Lambda=0.4$
and $\varepsilon_d=10.5$ and for the liquid
$\varepsilon_l=(1.33)^2$. Panel (a) shows the photonic band
structure for TM modes along the high-symmetry directions in the
1st Brillouin zone and panel (b) shows the corresponding photonic
density of states. The filled curve shows the density of states in
the absence of conduction while the superimposed curves are for
$\omega_D\Lambda/(2\pi c)=0.01$, 0.05, and 0.1. Panel (d) shows
the induced density of states in the center of the photonic band
gap. The data points correspond to the superimposed curves in
panel (b). } \label{fig1}
\end{figure}

In following we illustrate the above general results by numerical
simulations for a particular photonic crystal structure. For the
simulations of Eq.~(\ref{eq:E}) we employ a freely available
plane-wave method~\cite{Johnson:2001}. As an example we consider
the TM modes of a square lattice of high-index rods with diameter
$d/\Lambda=0.4$ and $\varepsilon_d=10.5$ and for the liquid
$\varepsilon_l=(1.33)^2$, see panel (c) in Fig.~\ref{fig1}. Panel
(a) shows the band structure and the corresponding density of
states is shown in panel (b) for different values of the Debye
frequency. The overall effect of the conductivity is obviously to
smear out sharp features as well as to introduce states in the
band-gap region of the unperturbed problem, see panel (d).

\begin{figure}[t!]
\begin{center}
\epsfig{file=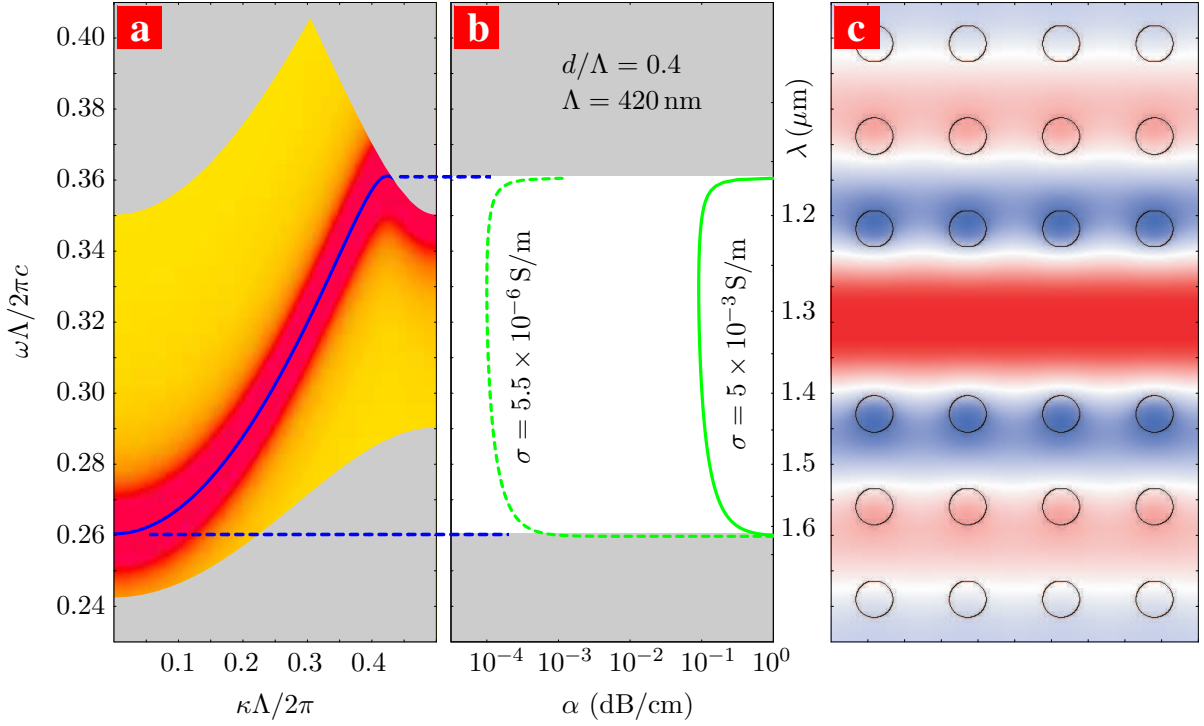, width=1\columnwidth,clip}
\end{center}
\caption{Liquid infiltrated line-defect photonic crystal waveguide
with a square lattice of high-index rods, see Panel (c), with
diameter $d/\Lambda=0.4$ and $\varepsilon_d=10.5$ while for the
liquid $\varepsilon_l=(1.33)^2$. Panel (a) shows the photonic band
structure (blue solid line) for propagation of TM polarized light
along the $\Gamma$X direction in the line-defect waveguide. The
color shading indicates the broadening of the line due to a finite
conductivity while the grey shading indicates the finite
density-of-states in the photonic crystal due to the projected
bands in the Brillouin zone. Panel (b) shows the Ohmic attenuation
for two different values of the conductivity corresponding to
highly purified water (dashed line) and typical drinking water
(solid line). The right $y$-axis shows the results in terms of the
free-space wavelength when results are scaled to a structure with
$\Lambda=420\,{\rm nm}$. Panel (c) shows the electrical field of
the waveguide mode at the $\Gamma$-point $\kappa=0$.} \label{fig2}
\end{figure}

In Fig.~\ref{fig2} we consider a line-defect waveguide structure
where a single line of rods has been removed. Panel (a) shows the
dispersion of the waveguide mode (blue solid line). As indicated
by the color shading, broadening of the mode is more pronounced
near the band edges compared to the central part of the band.
Panel (b) shows a corresponding increase in Ohmic attenuation near
the mode-band edges. For naturally occurring electrolytes the
attenuation will be of the order dB/cm in the center of the band
while highly purified water will result in an almost negligible
attenuation.

\section{Conclusion}

In conclusion we have studied the influence of Ohmic dissipation
on the attenuation and level broadening in liquid-infiltrated
photonic crystals. Our general results may readily be applied to
other types of liquid-infiltrated photonic crystals such as
three-dimensional opal structures as well as quasi two-dimensional
membrane structures. Our results illustrate that attention should
be paid to Ohmic dissipation and broadening for opto-fluidic
applications of photonic crystals in e.g. lab-on-a-chip systems.
One quite interesting consequence is the influence of the
electrolyte conductivity on the decay dynamics of excited
bio-molecules dissolved in the electrolyte. Similarly to the
observations for quantum dots in ordinary photonic
crystals~\cite{Lodahl:2004}, we expect that spontaneous emission
rates of bio-molecules can both be either suppressed (in band-gap
regions) or enhanced (e.g. by van Hove singularities) depending on
the lattice parameter and the photonic crystal design.
Potentially, the latter could be used to enhance inherently weak
signals of liquid-dissolved bio-molecules. Strong spectral
features in the density-of-states are a necessary condition for
this and the present work addresses the degree of persistence of
such features in the case of typical electrolytes.

\section*{Acknowledgments}

We thank Henrik Bruus for stimulating
discussions. This work is financially supported by the \emph{Danish Council for
Strategic Research} through the \emph{Strategic Program for Young
Researchers} (grant no: 2117-05-0037).


\newpage

\end{document}